\documentclass{article}

\usepackage[aux]{rerunfilecheck}
\usepackage{amsmath}
\usepackage{amssymb}
\usepackage{mathtools}
\usepackage[numbers,sort&compress]{natbib}
\usepackage[autostyle]{csquotes}
\usepackage{xfrac}
\usepackage{float}
\usepackage{graphicx}
\usepackage{color}
\usepackage{bbm} 
\usepackage{placeins} 
\usepackage{booktabs}

\title{\bf UV/IR Mixing, Causal Diamonds and the Electroweak Hierarchy Problem}

\author{Thomas W. Kephart$^1$ and Heinrich P\"as$^2$
\smallskip
\\
{\it $^1$ Department of Physics,
Vanderbilt University, Nahsville TN,
USA}\\
{\it $^2$ Fakult\"at f\"ur Physik,
Technische Universit\"at Dortmund,
Germany}
}

\begin{document}

\maketitle

\begin{abstract}
UV/IR Mixing is an umbrella term for phenomena in which high and low energy
physics does not decouple as expected and may offer new perspectives on the electroweak 
hierarchy problem, i.e. the apparent unnaturally large
hierarchy between the electroweak and the Planck scales. Based on how 
UV/IR mixing has been employed in the Cohen-Kaplan-Nelson bound and advocated as a solution to the cosmological constant problem, we argue that in the Higgs system
causal diamonds replace the cosmic horizon as an infrared bound for effective field theory and show how this ansatz may help to explain the large hierarchy between the electroweak and the Planck scale.
\end{abstract}

The electroweak hierarchy problem 
 is based on the observation that the contribution of Standard Model field one-loop corrections to the Higgs mass is quadratically divergent,
\begin{align}
\Delta m^2_H \sim \frac{\Lambda^2}{16 \pi^2}, 
\end{align}
where $\Lambda$ denotes the ultraviolet (UV) cutoff up to which the Standard Model is assumed to be valid
(for recent reviews see e.g. \cite{Craig:2022uua,Koren:2020pio}). 
If the Standard Model is understood as an effective field theory (EFT) valid up to e.g. the Planck scale, 
$\Lambda \sim M_{Pl}$,
this estimate implies that the Higgs particle should either have a mass expected to be of the order of the Planck scale, or that
finetuned cancellations among the various quantum corrections at different energy scales are extremely effective at conspiring to keep the Higgs mass light. Another way to think about this finetuning problem is the breakdown of naturalness, i.e. the feature that EFTs can be formulated  
in terms of local operators involving only light degrees of freedom (dofs), while the details of physics at high energy scales aren't correlated with the dynamics at large distances (the infrared or IR) and can be summarized into a finite number of parameters. 

Traditional approaches to address the electroweak hierarchy or fine tuning problem typically involve new physics beyond the Standard Model (BSM) at the TeV scale, that either would allow for a natural cancellation among the quantum corrections to the Higgs mass such as Supersymmetry, or lower the cutoff scale, such as Technicolor, Compositeness or Large Extra Dimensions. As yet no new physics beyond the Higgs particle has been found at the Large Hadron Collider, this gives rise to tight constraints on new particles and interactions in BSM theories, causing these theories to become increasingly fine tuned.

In this paper we pursue an alternative approach and reassess the question whether the expectation of a separation of scales is justified and under what conditions it may not be fulfilled. In doing so, we will adopt 
a popular example of UV/IR mixing that has received increased attention recently as a guiding principle, the 
Cohen-Kaplan-Nelson (CKN) bound on quantum corrections contributing to the dark energy density in the universe \cite{Cohen:1998zx}.
The argument is based on the well-known observation that quantum field theory (QFT) has a limited range of 
validity bounded by a UV cutoff $\Lambda$. An upper bound for this cutoff can be estimated as the energy
 beyond which quantum gravitational effects allow for the production of black 
holes. As these black holes will increase in size with higher energy, hence higher energies won't correspond to 
probing smaller distances, anymore. Cohen, Kaplan and Nelson then argue that this UV bound should be complemented with an IR cutoff originating from arguments about black hole entropy. In fact, according to
black hole thermodynamics,  the maximum information in a box of volume $L^3$ scales as the surface area of the box $\sim L^2$ instead of extensively $\sim L^3$ as would be expected in a conventional field theory. This implies that there must be a maximum length beyond which quantum field theories vastly overcount the degrees of freedom of the fundamental theory. CKN obtain an even more stringent bound for the IR cutoff $L_{IR}$ by requiring that the Schwarzschild radius 
\begin{align}
\frac{L_{IR}^3 \Lambda^4}{M_{Pl}^2}
\end{align}
of the maximum energy configuration $L_{IR}^3 \Lambda^4$ stored in the box should not exceed the size of the box $L_{IR}$, i.e. 
\begin{align}
L < L_{IR} \sim \frac{M_{Pl}}{\Lambda^2}.
\label{ckn}
\end{align}
This expression constitutes a relationship between the IR and UV cutoffs, according to which the
IR cutoff $L_{IR}$ scales with the inverse squared of the UV cutoff $\Lambda$.
CKN go on and adopt as the IR cutoff the present horizon size of the universe, obtaining a UV cutoff of $10^{-2.5}$~eV that is surprisingly close to the observed value of the cosmological constant. 

In the following we try to generalize the CKN argument and employ it to alleviate the large hierarchy between the 
electroweak and the SM cutoff scale. In doing so, at least two questions arise that should be addressed: First, it is obvious that the UV cutoff of $10^{2.5}$~eV, that protects the dark energy density from quantum corrections cannot be universal in the sense that it applies to every SM process. As it is well known that in general QFT works perfectly fine up to the highest scales tested so far, such a low energy UV cutoff would induce massive phenomenological problems. Second, the
argument of CKN is solely motivated by the question in which range QFT may safely ingnore any corrections from 
gravity. Apart from pointing out that there are cutoffs where QFT will not work anymore as expected, there is no physical reason given for where these cutoffs originate from. We thus will adopt that the cutoffs introduced by CKN are not universal, i.e. that different cutoffs apply for different physical processes, and offer some speculations about where these cutoffs may come from at the end of the paper.

As has been explained above, CKN used the correlation between IR and UV 
cutoffs obtained from an argument that the maximum length relevant for an experiment should be smaller than the
system's Schwarzschild horizon and combined that with a bound for the IR cutoff that they adopted to be the cosmic horizon to derive a reduced UV cutoff for quantum corrections. In the context of Feynman diagrams 
contributing to the Higgs mass, we argue that the relevant IR cutoff is provided by the
largest spacetime region that can be causally probed during the lifetime of the Higgs boson. 
In  \cite{Bousso:2000nf},
Raphael Bousso has
identified the maximum patch of spacetime that can be causally probed by a local observer and/or interfere
with a finite time quantum process as the causal diamond. Causal diamonds can be understood as the 
Minkowski space equivalent of horizons in curved or expanding spacetimes, compare e.g. \cite{Zurek:2022xzl} for a dictionary between black hole spacetimes and causal diamonds. A causal diamond
is defined as the intersection of an upward-opening lightcone
starting at the event where the process considered starts with a downward-opening lightcone ending at the event where the process stops.

We thus adopt the causal diamond constituted by the Higgs particle's lifetime $\Delta t$ as the relevant spacetime region defining the horizon scale for the IR bound. This gives rise to a length scale of the order of 
\begin{align}
L_{IR} \sim \Delta t \sim \left( 4~{\rm MeV} \right)^{-1}
\end{align}
 for the IR cutoff. Plugging this value $L_{IR}=\left(4~{\rm MeV}\right)^{-1}$ into eq.~(\ref{ckn}), we obtain a UV cutoff of
\begin{align}
\Lambda \sim 10^8~{\rm GeV}. 
\end{align}
While this value improved the amount of necessary fine tuning by 11 orders of magnitude, compared to  a Planck scale cutoff, it is still far from natural. 
(Another discussion of causal diamonds in the context of the CKN bound can be found in \cite{Banks:2019arz,2661076}.
Alternative interpretations  have been proposed that treat the CKN bound as a scale-dependent depletion of the QFT density of states
\cite{Banks:2019arz} or include the effects of physical particles \cite{Bramante:2019uub}.)

Yet, while this consideration obviously does not solve the hierarchy problem alone, it may at least contibute to a solution. 
One straightforward idea would be to complement UV/IR mixing from the causal diamond with a reduced Planck scale $M'_{Pl}<M_{Pl}$ as can be obtained in theories with large extra dimensions
\cite{Arkani-Hamed:1998jmv,Antoniadis:1998ig,Arkani-Hamed:1998sfv}. 
The recent PDG review on large extra dimensions reports the following bounds on the
fundamental quantum gravity scale \cite{PDG}:
Astrophysical processes such as supernova cooling, neutron star heating and contributions to the 
 diffuse gamma flux restrict $M’_P$ up to ${\cal O}(10^2-10^3)$~TeV, while collider experiments searching for signatures of missing transverse energy from gravitons escaping detection or black hole production
provide constraints of $M’_P>{\cal O}(1-10)$~TeV.
In our case, to solve the hierarchy problem,  it is sufficient to suppress the reduced Planck scale to $M'_P \sim 10^{9}$~GeV, still way above the TeV scale and safe from direct experimental testing at the LHC, to achieve $\Lambda \sim 2$~TeV and a natural Higgs mass
\begin{align}
m_H \sim \frac{1}{4 \pi} \Lambda \sim \frac{1}{4 \pi} \sqrt{\frac{M'_P}{L_{IR}}} \sim 10^2~{\rm GeV}.
\end{align}
Thus while neither large extra dimensions nor the CKN bound can resolve the electroweak hierarchy problem individually, combined these two effects actually can.

Finally we offer some speculation about where the new cutoffs introduced to protect QFT from gravitational effects 
may originate from. Before doing so, we would like to stress that our proposal is only one besides several other ideas 
that address the hierarchy problem in the context of UV/IR mixing. Many of these works explore models known as non-commutative field theories. Such theories work in a spacetime that is built up from tiny but finite cells and where the distance between two points depends on the order in which the distance in individual coordinates was measured. While these models  don’t describe any known physics yet, they can be motivated from string theory and seem to be promising approaches to address certain effects of quantum gravity.

In 2019, Nathaniel Craig and Seth Koren discussed how such toy models may help to figure out where the UV/IR mixing responsible for the light mass scale of the Higgs boson may originate from \cite{Craig:2019zbn}. One year later, Per Berglund, Tristan Hübsch and Djordje Minic studied a non-commutative version of string theory in which the mass of the Higgs boson is sensitive to both the UV and IR scales \cite{Berglund:2020qcu}. The concrete formula obtained resembles the one in the CKN paper. A different twist of UV/IR mixing in string theories had been pointed out by Steven Abel and Keith Dienes in 2021. String theories typically feature series of heavy copies of Standard Model particles, originating for example as standing wave excitations in curled up extra dimensions. As Abel and Dienes point out, many symmetries of string theory rely on the entire spectrum of these particles, including extremely heavy ones, and that these symmetries may be crucial to address the fine-tuning problems of dark energy and the Higgs boson \cite{Abel:2021tyt}.

Another effect of UV/IR mixing has been described to manifest itself as a 
a hidden symmetry that may reveal itself only when all individual Feynman diagrams contributing to an observable have been summed, or as a ``magic zero'' 
\cite{Arkani-Hamed-talk,Arkani-Hamed:2021xlp,Craig:2021ksw}. 
In other contexts, such hidden symmetries have been identified for example in the vanishing of the leading contribution to the anomalous dipole moments of the muon upon integrating out weak doublet and singlet vector-like fermions
that leads to an apparent UV-IR conspiracy at intermediate scales, and in turn have been advocated to shed 
new light on long-standing naturalness problems of the Standard Model and beyond .  
Yet,  while UV-IR mixing has been advocated as a candidate explanation for the  large electroweak hierarchy and other
fundamental fine-tuning problems in high energy physics, the phenomenon is usually 
considered as an exotic aspect of quantum gravity, or studied in contexts such as a hidden "Love" symmetry for black holes or non-commutative field theories, that have little obvious relevance to the electroweak hierarchy problem \cite{Berglund:2022qcc,Draper:2022pvk}. 

In the following we will argue that hidden symmetries or magic zeroes aren't necessarily an exotic phenomenon restricted to quantum gravity but may be a generic feature of entangled quantum systems. This conjecture 
may justify our study to generalize the CKN bound to non-gravitating systems. 
We start by considering the prototypical example of an entangled quantum system, the singlet Bell state
composed of two SU(2) doublet constituents:

\begin{align}
\psi = \frac{1}{\sqrt{2}}
\left( |\uparrow \downarrow \rangle - |\downarrow \uparrow \rangle \right).
\label{Bellsinglet}
\end{align}

The fact that $\psi$ is an SU(2) singlet implies that it respects a symmetry of the total system that is
not obvious from looking at the constituent SU(2) doublets. This is, in fact, the gist of the EPR paradox,
that in David Bohm's famous version considers the decay of an SU(2) singlet particle into two SU(2) doublets, so that 
these two constituents are perfectly anti-correlated in any measurement even if they are separated by a spacelike distance \cite{Bohm}. From the constituent viewpoint this (anti-)symmetry between degrees of freedom at different locations is indeed hidden or ``a magical zero''.

Now coming back to the entangled state, 
position is only one of infintely many possible Hilbert space bases. If we look at
Bohm's version of the EPR paradox for example in the momentum basis, we will observe that an equally surprising anti-correlation between the two different spins occurs if we label the spin states with momenta instead of position cordinates. It is thus natural to assume that entanglement creates surprising coincidences, not only between degrees of freedom at different locations, but also between degrees of freedom at different momenta or energies such as oscillation modes in a quantum field theory. 

Although the phenomenon so far is rarely discussed in the literature, it is well known that interacting quantum field theories give rise to momentum-space entanglement. In \cite{Balasubramanian:2011wt}, 
the reduced density matrix for the IR degrees of freedom and its relation to the conventional Wilsonian effective action has been derived.
More recently it has been advocated that renormalization, decoupling of heavy particle effects from low energy physics and the construction of effective field theories are intimately linked to momentum space entanglement
in QFTs \cite{Han:2020uwn}. In \cite{Klco:2021biu} such a UV-IR connection in EFTs has been explicitly
demonstrated by showing that the
entanglement between regions of the vacuum of a massless scalar field depends upon the UV-completion beyond a separation proportional to the UV cutoff scale of the theory (the relationship obtained between UV and IR differs from the one derived by CKN, though). Another potentially related study is the discussion of the Reeh-Schlieder theorem in 
\cite{Witten:2018zxz}. These works provide a strong hint that the argument above about UV/IR mixing from momentum-scale entanglement in quantum mechanics can be generalized to the framework
of interacting quantum field theories and that it is potentially relevant for the electroweak hierarchy problem. 
What’s more, entanglement may be closely related to gravity. In recent approaches by string theorists to derive space and time from quantum mechanics for example entanglement is directly related to distance in the emergent space
(see e.g. \cite{VanRaamsdonk:2010pw,Swingle:2009bg,Maldacena:2013xja}). 
There are works that may substantiate such associations. For example, in 2008 Gregory Minton and Vatche Sahakian found that a specific version of string theory features entanglement between physics at high and low energies
\cite{Minton:2007fd}. 
Also the vacuum of the a massless, non-interacting scalar field studied in  \cite{Klco:2021biu}
was found to dissolve into separable regions in space, with a size dependent on the cutoff energy, as a consequence of suppressed entanglement in the corresponding QFT. 

Next we address the question of how entanglement, which is usually introduced as a multi-particle phenomenon,
can be relevant for a one-particle system such as the Higgs boson. Such one-particle entanglement
has been discussed e.g. in \cite{enk2005,vedral2006}:
For example, the one-particle superposition
\begin{align}
\psi = \frac{1}{\sqrt{2}}
\left( |\uparrow \rangle - |\downarrow \rangle \right).
\end{align}
can be expressed in the occupation representation $|ij\rangle$ with $i$ representing the number of upwards pointing spins and $j$ specifying the number of downward  pointing spins as 
\begin{align}
\psi = \frac{1}{\sqrt{2}}
\left( |10 \rangle - |01 \rangle \right),
\end{align}
which exhibits the same form as the Bell singlet in eq.~(\ref{Bellsinglet}).
In a similar way, a quantum state $|ijk...\rangle$ can be employed to represent the various quantum corrections contributing to the Higgs mass at different energy or momentum scales. The superposition of the various 
Feynman diagrams then translates into one-particle entanglement 
of the corresponding quantum corrections. If some hidden symmetry between these entangled field modes
is effective in protecting the Higgs mass, it will be broken by decoherence, i.e. the tracing out
of dofs that are unaccessible from the perspective of a local observer. As the Higgs is the only fundamental scalar 
we know of, it may be a specific feature of scalar fields to possess such hidden symmetries.
Moreover, since it is the Higgs that sets the mass scale of all the
SM particles, except for maybe the neutrinos, we believe
the Higgs causal diamond is more fundamental than the
causal diamond of any of the other SM particles. Indeed,
if the EW symmetry was not broken by the Higgs potential,
then all three families of fermions would be massless and 
not decay at all.

Of course, the Higgs mass is small compared to the expectation (the Planck scale) yet non-vanishing.
Thus in the context of the electroweak hierarchy problem any hidden symmetry implied by momentum-space entanglement has to be broken. Translating this problem back to the context of quantum mechanics, such symmetry breaking can be realized via decoherence. If only a small number of degrees of freedom constributing to the observable considered in this context is traced out, the
original symmetry will be only slightly broken, giving rise e.g. to a non-zero but small observable that is natural
in t'Hooft's sense. Indeed, as has been emphasized in \cite{Klco:2021lap}, entanglement entropy can serve as an
order parameter for symmetry breaking. 

While the CKN argument  relies on the notion of horizons, in a quantum context such horizons would determine which dofs have to be traced over leading to a decohered, mixed state or reduced density matrix for the remainder of the system \cite{Zeh:1970zz}.
Indeed, the horizon of the causal diamond then defines an observer-independent choice of environment for decoherence. Since the causal diamond's lightlike boundary constitutes a one-way membrane for information, all dofs that leave the diamond are lost with respect to the finite time experiment defining the
diamond, and thus have to be traced over
\cite{Bousso:2011up}. 
The smaller a causal diamond for a concrete process is, the more dofs of the total entangled system, i.e. 
the entire quantum universe, arguably the complete spacetime created e.g. in inflation including areas beyond our cosmic horizon, have to be traced out. Thus, qualitatively speaking, a large horizon such as the cosmic horizon employed in the CKN bound corresponds to a smaller number of decohered dofs and thus a marginal
symmetry breaking effect (such as the tiny dark energy density in the universe), while a small causal diamond 
corresponds to a large number of dofs traced out and thus a significant symmetry breaking effect.

If we deal with a decaying particle such as the Higgs boson, then
the causal diamond will be defined by the particle's  lifetime.
In momentum space, this will translate into an IR cutoff for low-energy oscillation modes.
Such ``open EFTs'' obtained by tracing out momentum states have also been discussed in the different context 
of cosmology, for quantum fluctuations
of the inflaton field arising in cosmic inflation, see e.g. 
\cite{Burgess:2015ajz,Kaplanek:2019dqu,Kaplanek:2019vzj,Brahma:2022yxu}.

In summary, we proposed a generalization of the CKN mechanism to protect the dark energy density in the universe
from diverging quantum corrections to the electroweak hierarchy problem. We showed that by replacing the 
Hubble horizon with the causal diamond of the Higgs particle, the hierarchy between the electroweak and the Planck scale can be ameliorated and even resolved, if in addition extra dimensions of spacetime are introduced.
The scenario suggests that new physics and/or quantum gravity phenomina may arise at a scale of $10^5$~TeV, well above 
the energies probed at present by the LHC, but not so far away that it is entirely inconceivable that future experiments
searching for example for precision tests of quantum gravity at the highest possible energies may be sensitive 
to these effects.       
Apart from the difficulties of testing this phenomenon in particle physics, there  may arise  new possibilities  
to study the phenomenon experimentally in quantum systems in condensed matter, or in quantum optics,
or in quantum simulations of QFTs.


\end{document}